
\newbox\leftpage
\newdimen\fullhsize
\newdimen\hstitle
\newdimen\hsbody
\tolerance=1000\hfuzz=2pt

\def\bigans{b }
\def\answ{b }
\ifx\answ\bigans\message{(this will come out unreduced.}
\magnification=1200\baselineskip=20pt
\font\titlefnt=amr10 scaled\magstep3\global\let\absfnt=\tenrm
\font\titlemfnt=ammi10 scaled\magstep3\global\let\absmfnt=\teni
\font\titlesfnt=amsy10 scaled\magstep3\global\let\abssfnt=\tensy
\hsbody=\hsize \hstitle=\hsize 

\else\def\apans{h }
\message{(this will be reduced.}
\let\lr=l
\magnification=1000\baselineskip=16pt\voffset=-.31truein
\hstitle=8truein\hsbody=4.75truein\vsize=7truein\fullhsize=10truein
\ifx\apansw\apans\special{ps: landscape}\hoffset=-.54truein
  \else\hoffset=.05truein\fi
\font\titlefnt=amr10 scaled\magstep4 \font\absfnt=amr10 scaled\magstep1
\font\titlemfnt=ammi10 scaled \magstep4\font\absmfnt=ammi10 scaled\magstep1
\font\titlesfnt=amsy10 scaled \magstep4\font\abssfnt=amsy10 scaled\magstep1

\output={\ifnum\count0=1 
  \shipout\vbox{\hbox to \fullhsize{\hfill\pagebody\hfill}}\advancepageno
  \else
  \almostshipout{\leftline{\vbox{\pagebody\makefootline}}}\advancepageno
  \fi}
\def\almostshipout#1{\if l\lr \count1=1
      \global\setbox\leftpage=#1 \global\let\lr=r
   \else \count1=2
      \shipout\vbox{\hbox to\fullhsize{\box\leftpage\hfil#1}}
      \global\let\lr=l\fi}
\fi
\def\pmb#1{\setbox0=\hbox{#1}%
 \kern-.025em\copy0\kern-\wd0
 \kern .05em\copy0\kern-\wd0
 \kern-.025em\raise.0433em\box0 }

\def\title#1#2{\nopagenumbers\absfnt\hsize=\hstitle\rightline{}%
\centerline{\titlefnt\textfont0=\titlefnt%
\textfont1=\titlemfnt\textfont2=\titlesfnt #1}%
\centerline{\titlefnt\textfont0=\titlefnt%
\textfont1=\titlemfnt\textfont2=\titlesfnt #2
}%
\textfont0=\absfnt\textfont1=\absmfnt\textfont2=\abssfnt\vskip .5in}

\def\date#1{\vfill\leftline{#1}%
\tenrm\textfont0=\tenrm\textfont1=\teni\textfont2=\tensy%
\supereject\global\hsize=\hsbody%
\footline={\hss\tenrm\folio\hss}}
%

\def\nolabels{\def\eqnlabel##1{}\def\eqlabel##1{}\def\reflabel##1{}}
\def\writelabels{\def\eqnlabel##1{\hfill\rlap{\hskip.09in\string##1}}%
\def\eqlabel##1{\rlap{\hskip.09in\string##1}}%
\def\reflabel##1{\noexpand\llap{\string\string\string##1\hskip.31in}}}
\nolabels
%
\global\newcount\secno \global\secno=0
\global\newcount\meqno \global\meqno=1

\def\newsec#1{\global\advance\secno by1\xdef\secsym{\the\secno.}\global\meqno=1
\bigbreak\bigskip
\noindent{\bf\the\secno. #1}\par\nobreak\medskip\nobreak}
\xdef\secsym{}

\def\appendix#1#2{\global\meqno=1\xdef\secsym{#1.}\bigbreak\bigskip
\noindent{\bf Appendix #1. #2}\par\nobreak\medskip\nobreak}


\def\eqnn#1{\xdef #1{(\secsym\the\meqno)}%
\global\advance\meqno by1\eqnlabel#1}
\def\eqna#1{\xdef #1##1{(\secsym\the\meqno##1)}%
\global\advance\meqno by1\eqnlabel{#1$\{\}$}}
\def\eqn#1#2{\xdef #1{(\secsym\the\meqno)}\global\advance\meqno by1%
$$#2\eqno#1\eqlabel#1$$}

\global\newcount\ftno \global\ftno=1
\def\refsymbol{\ifcase\ftno
\or\dagger\or\ddagger\or\P\or\S\or\#\or @\or\ast\or\$\or\flat\or\natural
\or\sharp\or\forall
\or\oplus\or\ominus\or\otimes\or\oslash\or\amalg\or\diamond\or\triangle
\or a\or b \or c\or d\or e\or f\or g\or h\or i\or i\or j\or k\or l
\or m\or n\or p\or q\or s\or t\or u\or v\or w\or x \or y\or z\fi}
\def\foot#1{{\baselineskip=14pt\footnote{$^{\refsymbol}$}{#1}}\ %
\global\advance\ftno by1}


\global\newcount\refno \global\refno=1
\newwrite\rfile
\def\ref#1#2{$^{(\the\refno)}$\nref#1{#2}}
\def\nref#1#2{\xdef#1{$^{(\the\refno)}$}%
\ifnum\refno=1\immediate\openout\rfile=refs.tmp\fi%
\immediate\write\rfile{\noexpand\item{\the\refno.\ }\reflabel{#1}#2.}%
\global\advance\refno by1}
\def\addref#1{\immediate\write\rfile{\noexpand\item{}#1}}

\def\semi{;\hfil\noexpand\break}



\hyphenation{anom-aly anom-alies coun-ter-term coun-ter-terms}

 \def\bb{{\bf  b}}



\def\bJ{{\bf J}}
\def\bGam{\pmb{$\Gamma$}}
\def\bomg{\pmb{$\omega$}}

\centerline{{\bf Two-dimensional Laplacian growth}}

\centerline{{\bf can be mapped onto Hamiltonian dynamics}}

\bigskip
\centerline{Raphael Blumenfeld\foot{Current address: Center for Nonlinear
studies, MS B258, Los Alamos National Laboratory, Los Alamos, NM 87545}}

\centerline{Princeton Materials Institute, Princeton University}
\centerline{Bowen Hall, 70 Prospect Avenue, Princeton, NJ 08540-5211, USA}

\bigskip
\item{}{\bf Abstract}
\smallskip
It is shown that the dynamics of the growth of a two dimensional surface in a
Laplacian field can be mapped onto Hamiltonian dynamics. The mapping is carried
out in two stages: first the surface is conformally mapped onto the unit
circle, generating a set of singularities. Then the dynamics of these
singularities are transformed to Hamiltonian action-angle variables. An
explicit condition is given for the existence of the transformation. This
formalism is illustrated by solving explicitly for a particular case where the
result is a separable and integrable Hamiltonian.

\bigskip
PACS numbers: 68.70+w, 81.10.Dn, 11.30.Na

\date{LA-UR-93 - 2800}
\eject
\bigskip
\smallskip
Much effort has been directed in recent years towards understanding Laplacian
growths, both due to the rich variety of patterns that they display and because
of their occurence in many natural and man-made systems. Paradigmatic cases are
diffusion-limited aggregation, solidification of supercooled liquid and
electrodeposition.\ref\rev{For review see, e.g., P. Pelce, "Dynamics of Curved
Fronts" (Academic Press, San Diego, 1988); D. A. Kessler, J. Koplik and H.
Levine, Adv. Phys. {\bf 37}, 255 (1988); P. Meakin, in "Phase Transitions and
Critical Phenomena" Vol. 12 (Academic Press, 1988) Eds. C. Domb and J. L.
Lebowitz; T. Vicsek, "Fractal Growth Phenomena" (World Scientific, Singapore,
1989)} In spite of more than a decade of intensive research into the problem
there is still no theory that can predict the statistics of the patterns that
such processes lead to. These moving-boundary problems are deceptively easy to
formulate yet very difficult to solve analytically. It has been
proposed\ref\galin{L. A. Galin, Dokl. Akad. Nauk USSR {\bf 47}, 246 (1945); P.
Ya. Polubarinova-Kochina, Dokl. Akad. Nauk USSR {\bf 47}, 254 (1945); Prikl.
Math. Mech. {\bf 9}, 79 (1945)}$^,$\ref\rich{S. Richardson, J. Fluid Mech.,
{\bf 56}, 609 (1972)} to conformally map the physical surface of such a growth
in two dimensions onto the unit circle (UC) and study the evolution of the
singularities of the map.\ref\sb{B. Shraiman and D. Bensimon, Phys. Rev. {\bf A
30}, 2840 (1984)} The resulting equations of motion (EOM's) are strongly
coupled nonlinear first order ODE's that are difficult to solve, other than for
special cases.\ref\all{L. Paterson, J. Fluid Mech. {\bf 113}, 513 (1981); L.
Paterson, Phys. Fluids {\bf 28}, 26 (1985); S. D. Howison,  J. Fluid Mech. {\bf
167}, 439 (1986); D. Bensimon and P. Pelce, Phys. Rev. {\bf A33}, 4477 (1986);
S. Sarkar and M. Jensen, Phys. Rev {\bf A35}, 1877 (1987); B. Derrida and V.
Hakim, Phys. Rev. {\bf A45}, 8759 (1992)} Another difficulty with this approach
is that the formalism breaks down after a finite time because singularities of
the conformal map travel to the UC and eventually hit it, at which time the map
ceases to be analytic. This breakdown is manifested in cusp singularities that
form on the physical surface in the absence of surface tension.\ref\leo{D.
Bensimon, L. P. Kadanoff, S. Liang, B. I. Shraiman and C. Tang, Rev. Mod. Phys.
{\bf 58}, 977 (1986); W-s Dai, L. P. Kadanoff and S. Zhou, Phys. Rev. {\bf A
43}, 6672 (1991)} A very important question in this approach is whether the
system is integrable or even Hamiltonian. It has been found that the problem
enjoys a set of conserved quantities,\rich$^,$\ref\mm{M. B. Mineev-Weinstein,
Physica {\bf D 43}, 288 (1990); M. B. Mineev-Weinstein and S. P. Dawson,
preprint (1993)} but it is unclear how these quantities can assist in finding
an energy-like functional in the problem.

In this paper it is shown that it is possible to transform the dynamics that
govern the growth of the surface into Hamiltonian dynamics as long as the EOM's
hold (namely, up to the cusp formation). This is done in two steps: First the
EOM's of the singularities of a general map are written down and then their
space-coordinates are transformed into action-angle variables. The equations
for this transformation are given for any general initial surface. Although the
general existence of a solution for these equations is not rigorously proven
here, I show that the surface evolves according to Hamilton's equations, and an
example is given where a particular system is solved exactly and is shown to be
integrable for a family of arbitrary initial conditions. The action variables
are constants of the motion and an energy functional can be clearly identified.
The importance of such a mapping cannot be overemphasized: i) By producing such
a transformation the system is demonstrated to be Hamiltonian and possibly also
integrable, which is an important issue in itself; ii) A large body of
knowledge that has been accumulated through many decades of studying
Hamiltonian systems is consequently made accessible to such growth problems.
This can immediately pave the way to a theory of Laplacian growth, as will be
reported elsewhere.\ref\bi{R. Blumenfeld, in preparation}
\bigskip

I start by formulating the problem and briefly presenting the EOM's of the
singularities that are generated by the conformal map of the surface onto the
UC. Consider a simply connected line surface, $\gamma(s)$, embedded in two
dimensions which is parametrized by $0\leq s <2\pi$, and which is fixed at a
given electrostatic potential. A higher potential is assigned to a circular
boundary whose radius tends to infinity. The potential field, $\Phi$, outside
the area enclosed by $\gamma$ is determined by Laplace's equation
\eqn\L{\nabla^2 \Phi = 0\ .}
The surface is assumed to grow at a rate that is proportional to the local
electrostatic field normal to the surface
$$v_n = - {\bf \nabla}\Phi\cdot\hat{\bf n}\ .$$

Shraiman and Bensimon\sb have shown that the surface evolves in time, $t$,
according to
\eqn\Ai{\partial_t\gamma(s,t) = -i\partial_s\gamma(s,t)
\Bigl[|\partial_s\gamma(s,t)|^{-2} +
ig\Bigl(|\partial_s\gamma(s,t)|^{-2}\Bigr)\Bigr]\ ,}
where $g$ is a real function that corresponds to a physically insignificant
'slide' of a point along the surface. Without that last term the EOM can be
written as
\eqn\Dh{\partial_t\gamma(s,t) = -i{{\delta s}\over{\delta \gamma^*}}\ ,}
where $^*$ stands for complex conjugate. Eq. $\Ai$ is the limit of a conformal
map $\zeta=F(z)$ that maps the UC onto the physical surface in the
$\zeta$-plane through $\gamma(s,t)=\lim_{z\to e^{is}} F(z,t)$. The map
considered here is a general ratio of two polynomials of the same degree. This
requirement results from the need that the topology far away from the growth
remain unchanged, so as to retain the original boundary conditions on a circle
far away and the generic logarithmic divergence of $\Phi$ when
$z\to\infty$.\ref\bb{R. Blumenfeld and R. C. Ball, in preparation}
\eqn\Aii{ F' \equiv {{dF(z)}\over{dz}} = \prod_{n=1}^N {{z-Z_n}\over{z-P_n}}\
,}
where $\{Z\}$ and $\{P\}$ are the zeros and the poles of the map, respectively.
It can be shown\sb$^,$\bb that the dynamics of these singularities are governed
by the EOM's
\eqn\Aiii{\eqalign{-\dot Z_n &= Z_n\Bigl\{G_0 + \sum_{m'} {{Q_n +
Q_{m'}}\over{Z_n - Z_{m'}}} \Bigr\} + Q_n\Bigl\{1 - \sum_m{{Z_n}\over{Z_n -
P_m}}\Bigr\} \equiv f_n(\{Z\};\{P\}) \cr
-\dot P_n &= P_n\Bigl\{G_0 + \sum_m {{Q_m}\over{P_n - Z_m}}\Bigr\} \equiv
g_n(\{Z\};\{P\})\ , \cr }}
where
$$\eqalign{Q_n &= 2\prod_{m=1}^N\Bigl[ {{(1/Z_n - P_m^*)(Z_n-P_m)}\over
{(1/Z_n - Z_m^*)(Z_n - Z_{m'})}}\Bigr] \cr
G_0 &= \sum_{m=1}^N {{Q_m}\over{2 Z_m}} + \prod_{m=1}^N {{P_m}\over{Z_m}}\ ,
\cr}$$
and where the primed index indicates $m'\neq n$.

We now wish to introduce a new set of canonical coordinates that can transform
this system of coupled first order ODE's into a Hamilton system. There are two
reasons why one should expect to find a Hamiltonian at all: The first stems
from the existence of constants of the motion, as found by Richardson.\rich The
second reason relates to the fact that already the EOM of the surface can be
written in the form of Hamilton's equations as follows: Eq. $\Ai$ is the limit
of an EOM for the map $F$ as $z\to e^{is}$:
\eqn\Aia{ {{\partial F'}\over{\partial t}} = {\partial\over{\partial z}} \bigl[
z F' G\bigr]\ ,}
where $G$ is the analytic function whose limit is the square brackets on the
r.h.s. of $\Ai$. Using the identity
$$\Biggl({{\partial F'}\over{\partial t}}\Biggr) \Biggl({{\partial
t}\over{\partial z}}\Biggr) \Biggl({{\partial z}\over{\partial F'}}\Biggr) =
-1$$
we obtain that
\eqn\Aib{{{\partial z}\over{\partial t}} = -{{\partial (z F' G)}\over{\partial
F'}}\ .}
equations $\Aia$ and $\Aib$ can be interpreted then as Hamilton's relations if
$z$ and $F'$ are interpreted as canonical field variables and $z F' G$ as the
'Hamiltonian density'. Alternatively, Eq. $\Dh$ immediately shows the
underlying Hamiltonian structure of the EOM.

Since the dynamical system $\Aiii$ represents the motion of the curve it is
plausible that this system can also be derived from a Hamiltonian. To this end
let us start from the desired result and work our way backwards. The goal is to
obtain a Hamiltonian of separable variables (although separability is not
necessarily a constraint)
\eqn\Aiv{H=H(\{\bJ\};\{{\bf \Theta}\})\ \ \ ;\ \ \ \bJ\equiv(J_1, J_2,\dots,
J_N)\ ,\ \ \ {\bf \Theta}\equiv(\Theta_1, \Theta_2,\dots, \Theta_N)\ .}
If this system is integrable then $H=H\{\bJ\}$ and we have a motion on a
$N$-dimensional torus. At this stage it is possible to choose arbitrarily the
form of the target Hamiltonian into which the system should be mapped, but for
simplicity we require here that the Hamiltonian have the form
\eqn\Av{H=\sum_{n=1}^N \omega_n J_n\ .}
It should be emphasized though that the formulation presented here is in no way
restricted to this form. These coordinates are required to obey Hamilton-Jacobi
EOM's
\eqn\Avi{\dot\Theta_n = {{\partial H}\over{\partial J_n}}\ \ \ ;\ \ \ \dot{J_n}
= -{{\partial H}\over{\partial \Theta_n}}\ .}
The transformation we seek should yield $J_n=J_n(\{Z\};\{P\})$ and
$\Theta_n=\Theta_n(\{Z\};\{P\})$. Combining equations $\Aiii$, $\Av$ and $\Avi$
we obtain that to effect the desired transformation the following set of
equations needs to be satisfied
\eqn\Avii{\dot J_n = \sum_{m=1}^N \Bigl[{{\partial J_n}\over{\partial Z_m}}f_m
+ {{\partial J_n}\over{\partial P_m}}g_m \Bigr] = -{{\partial H}\over{\partial
\Theta_n}} = 0 }
\eqn\Aviii{\dot \Theta_n = \sum_{m=1}^N \Bigl[{{\partial
\Theta_n}\over{\partial Z_m}}f_m + {{\partial \Theta_n}\over{\partial P_m}}g_m
\Bigr] = {{\partial H}\over{\partial J_n}} = \omega_n \ ,}
where the r.h.s. of these equations is particular to the Hamiltonian that we
have chosen in Eq. $\Av$. This set of equations can be written generally for
any system in the form:
\eqn\Adi{\bigl({\bf f}(\bGam)\cdot{\bf \nabla}\bigr) \bJ(\bGam) = \bomg \ .}
In this notation ${\bf f}$, $\bJ$, $\bGam$ and $\bomg$ are $2N$-components
vectors
$$\eqalign{ {\bf f} &\equiv \bigl(f_1, f_2,\dots, f_N, g_1, \dots, g_N \bigr) \
\ ; \ \ \ \bGam \equiv \bigl(Z_1, Z_2,\dots, Z_N, P_1, \dots, P_N \bigr) \cr
{\bJ} &\equiv \bigl(J_1, J_2,\dots, J_N, \Theta_1, \dots, \Theta_N \bigr) \ \ ;
\ \bomg \equiv \bigl(-{{\partial H}\over{\partial \Theta_1}}, -{{\partial
H}\over{\partial \Theta_2}},\dots, -{{\partial H}\over{\partial \Theta_N}},
{{\partial H}\over{\partial J_1}}, \dots, {{\partial H}\over{\partial J_N}}
\bigr) \ , \cr}$$
and ${\bf \nabla}$ is the gradient in the $2N$-dimensional space of $\bGam$. If
$\bomg$ is a vector of $N$ zeros and $N$ constants of the motion, as chosen in
Eq. $\Av$ then $\Adi$ is a linear set and therefore has a solution as long as
the operator ${\bf f}(\bGam)\cdot{\bf \nabla}$ has no vanishing eigenvalue.
Thus, unless such a singular case occurs, this set of equations does have a
solution for $\{\bJ(\bGam)\}$ and $\{{\bf \Theta(\bGam)}\}$. This solution,
when it exists, defines a specific transformation from the chosen Hamiltonian
to the problem of the dynamics of the singularities, and hence to the original
growth problem. The existence of such a solution immediately points to the
integrability of the system.
\bigskip
Having discussed the general case, it is useful to consider a specific example
of a class of initial conditions where the solution to the set of equations
$\Avii$ and $\Aviii$ indeed exists and can be obtained explicitly. Assume that
at $t=0$ the initial surface can be represented by
\eqn\Bi{\gamma(s,0) = e^{is} - \sum_{j=1}^2 R_j\ln\bigl(e^{is}-P_j(0)\bigr)\ ,}
where $R_j = (-1)^j [(P_j^2(0) - Z_1^2(0))/(P_1(0)-P_2(0))]$ and $|P_1|, |P_2|,
|Z_1| < 1$. The surface $\gamma(s,t)$ can be shown to retain this form for any
later time by substituting for $P_j$ and $Z_j$ their time-dependent values.
This form is valid for any number of singularities when $-R_j$ is replaced by
the residues of the analytic map $F$. The growth problem consists now of
finding the dynamics of two zeros at $Z_1(t)$ and $Z_2(t)=-Z_1(t)$ and two
poles at $P_1(t)$ and $P_2(t)$. The trajectories of these singularities can be
found by substituting into $\Aiii$:
\eqn\Biii{\eqalign{-\dot Z_1 &= Z_1 \Bigl[{{2Q_1}\over {Z_1}} -
{Q_1\over{Z_1-P_1}} - {Q_1\over{Z_1-P_2}} - {{P_1 P_2}\over{Z_1^2}}\Bigr] \cr
-\dot P_j &= P_j \Bigl[{{Q_1}\over{Z_1}} + {{Q_1}\over{P_j-Z_1}} -
{{Q_1}\over{P_j+Z_1}} - {{P_1 P_2}\over{Z_1^2}}\Bigr] \ ,\cr}}
It can be shown\bb that for the map to be analytic this 2x2 system has to obey
the relation $Z_1+Z_2=P_1+P_2$. This immediately simplifies the EOM's to the
form
\eqn\Biiia{\eqalign{{d\over{dt}} Z_1^2 &= 2P_1 P_2 \bigl[1 - 2K\bigr] \cr
{d\over{dt}} P_j &= {P_j\over Z_1^2} \Bigl[ P_1 P_2 + (P_j^2 + Z_1^2) K \Bigr]
\ ,\cr} }
where $K\equiv (1-Z_1P_1^*)(1-Z_1P_2^*)/(1-|Z_1|^4)$. I choose the initial
conditions such that $|P_j(0)|$ is smaller than $|Z_1(0)|$ and also
arg$(|P_j|={\rm arg}(|Z_j|)$. This choice is inconsequential to the exact
solution below and for the purpose of the present discussion. For the opposite
choice,  $|P_j(0)|>|Z_1(0)|$, the growth process is stable (namely, no cusp
will form) and the solution holds to $t\to\infty$. The unstable case is chosen
to emphasize the thrust of this paper that the system is integrable until the
EOM's loose their validity regardless of whether this occurs at a finite or
infinite time.

A rather tedious manipulation of equations $\Biiia$ yields that the following
is a constant of the motion
\eqn\Biiib{J_j = {1\over{Z^2_j-P^2_j}} + \zeta_j = const.\ ,}
where $\zeta_j = {1\over 8}\ln{\Bigl[{{1+P^2_j}\over{1-P^2_j}}\Bigr]}$. These
constants we can imediately identify as the action variables. By substituting
for $Z_1^2$ from $\Biiib$ and employing some algebra the solution for $P^2_j$
is found
\eqn\Biiic{t = {1\over 2}\ln{\Bigl[P^2_j(J_j-\zeta_j)^8\Bigr]} - \int^{\zeta_j}
{{2 d\zeta}\over{{\rm tgh}4\zeta\ (J_j-\zeta)^2}}\ .}
Using relations $\Biiib$ and $\Biiic$ we obtain the solution for $Z_1$ in a
similar form. From the EOM's it can be verified that from the above initial
conditions the zeros propagate towards the UC faster than the poles, and hence
the solution exists as long as the singularities are within the UC. Several
stages of the growth are shown in Fig. 1, where the surface evolves from mildly
oval into an eye-shaped structure. In the mathematical plane the evolution
consists of the poles and the zeros moving radially towards the UC.

It should be noted that the above treatment assumes a unity factor in front of
the map $F$. In fact there should also appear a purely time-dependent prefactor
which takes care of the total area of the growth increasing linearly with time
(The assumption is that the flux into the growth is constant in time). By
setting this prefactor to unity the surface is effectively rescaled at each
time step, which is why in Fig. 1 sections of the boundary seem to retreat with
time. The evolution of this prefactor can be very simply incorporated into the
formulation, but for clarity it has been omitted here.
\smallskip
Since we have now solutions in the form of Eq. $\Biiic$
$$t = I_{\Gamma_j}\ \ ; \ \ \ \bGam \equiv \bigl(Z_1, Z_2,\dots, Z_N, P_1,
\dots, P_N \bigr)$$
we can use Eqs. $\Avii$ and $\Aviii$ to find the action-angle variables in
terms of the original coordinates. From the integrals of motion we obtain
\eqn\Biv{\Biggl\{\matrix {J_k \cr \Theta_k}\Biggr\} = \sum_{j=1}^2 \Biggl\{
\matrix {a_{jk} \cr c_{jk}}\Biggr\} I_{Z_j} + \Biggl\{\matrix {b_{jk} \cr
d_{jk}} \Biggr\} I_{P_j} \ ,}
where $a, b, c$, and $d$ are constants. These coefficients of the action-angle
variables are required to obey
$$\sum_{j=1}^2 \Biggl\{\matrix {a_{jk} + b_{jk} \cr c_{jk} + d_{jk} }\Biggr\} =
\Biggl\{\matrix {0 \cr \omega_j }\Biggr\}\ .$$

It may seem from this result that there are too many coefficients that can be
chosen arbitrarily. But for a general growth these coefficients are complex and
the requirement that the action and angle variables take on only real values
reduces the arbitrariness to exactly four independent coefficients with
possible other four arbitrary integers (corresponding to multiplicity of $2\pi
i$ when setting the imaginary parts of the variables to zero). The angle
variables, $\Theta_k$, thus grow linearly with time while the action variables,
$J_k$, are constants of the motion, and by the transformation $\Biv$ we have
obtained the target Hamiltonian $\Av$.
\bigskip
To conclude I have shown here that the growth of a free surface in a Laplacian
field is governed by Hamiltonian dynamics which is integrable if $\Adi$ has a
solution. This formalism holds until cusp singularities occur. I have shown
that the surface evolves according to Hamilton's equations if $F'$ and $z$ are
interpreted as the canonical field variables and the Hamiltonian density is
$zF'G$. A family of initial conditions has been analysed explicitly where
integrability can be demonstrated, showing that at least for some classes of
arbitrary initial conditions relation $\Adi$ does have a solution. I have
chosen the simple Hamiltonian given in $\Av$ to illustrate how this
transformation can be carried out, and have formulated the transformation
equations for a general surface. Several questions still remain: i) Is there a
case where the operator ${\bf f}(\bGam)\cdot{\bf \nabla}$ in Eq. $\Adi$ has at
least one vanishing eigenvalue? and if so what is the structure it corresponds
to? ii) How constrained are we in choosing the Hamiltonian so that the
transformation equations still have a solution? These and other questions are
currently under investigation. This author believes that the example given here
represents only a 'Hydrogen Model' system where direct mapping to Hamiltonian
dynamics is possible, and that such a mapping is a general property of
Laplacian growth processes. Therefore this approach can pave the way towards a
theory of Laplacian (and possibly more general) growth processes, as will be
discussed in a later publication that is currently under preparation.

\bigskip
{\bf Acknowledgement}

It is a pleasure to thank Robin C. Ball for stimulating discussions, Gennady
Berman for suggesting the action-angle separation as in $\Av$, and Darryl D.
Holm for critical reading and helpful suggestions. This work began at Princeton
Materials Institute, Princeton University and was completed at the Center for
Nonlinear Studies, Los Alamos National Laboratory. This work is supported in
part by grant No. DE-FG02-92ER14275 of U.S. Department of Energy.

\bigskip
{\bf FIGURE CAPTIONS}
\bigskip
\item{1.} The growth of the surface discussed in the text for two zeros
$Z^2_j(t_0)=0.55$ and two poles $P^2_j(t_0)=0.005$.
\vfill\eject\immediate\closeout\rfile
\baselineskip=18pt\centerline{{\bf REFERENCES}}\bigskip\frenchspacing%
\input refs.tmp\vfill\eject\nonfrenchspacing
\bye